\documentstyle[11pt,newpasp,twoside,epsf,psfig]{article}
\def\edcomment#1{\iffalse\marginpar{\raggedright\sl#1\/}\else\relax\fi}
\reversemarginpar

\begin{document}

\title{Pre-main sequence binaries with aligned disks ?}
\author{
  Sebastian Wolf$^{1}$, Bringfried Stecklum$^{1}$, and Thomas Henning$^{2}$}
\affil{$^{1}$Th\"uringer Landessternwarte Tautenburg, 
  Sternwarte 5, D--07778 Tautenburg, Germany}
\affil{$^{2}$Astrophysikalisches Institut und Universit\"ats-Sternwarte, 
  Schillerg\"asschen 2-3, D--07745 Jena, Germany}

%-----------------------------------------------------------------------------------------
\begin{abstract}
We present the results of a study performed with the goal to investigate whether low-mass 
pre-main sequence binary stars are formed by multiple fragmentation
or via stellar capture. If binaries form preferentially by fragmentation,
we expect their disks to be co-planar. On the other hand, the capture scenario
will lead to a random distribution of disk orientations.
We performed near-infrared polarization measurements of 49 young close binary stars 
in the K band with SOFI at the NTT. The near-infrared excess radiation
of the targets mostly point to the presence of disks. For a major fraction of 
the sample, evidence for disks is also obvious from other features (outflows, 
jets, Herbig-Haro objects). We derived the disk orientation from the orientation
of the polarization vector of both components of each binary.
This statistical study allows to test which hypothesis (co-planarity, random
orientation) is consistent with the observed distribution of polarimetric
position angles. We find evidence that the disks are preferentially aligned.
\end{abstract}


%-----------------------------------------------------------------------------------------
\section{Introduction}\label{intro}

During recent years, near-infrared (NIR) imaging and high-resolution 
observations yielded conclusive evidence that most (if not all) low-mass
stars are born in binary and multiple stellar systems 
(Simon et al.\ 1995, Ghez et al.\ 1997,
Leinert et al.\ 1997, K\"ohler \& Leinert 1998). This finding suggests that binary
formation is the rule and the birth of single stars the exception.
However, the formation and survival 
of binary systems obviously depend upon environmental conditions 
(Bouvier et al.\ 1997).
Up to now, it is not known at which stage the binary/multiple star formation
mode becomes dominant. Numerical studies suggest that this occurs at early
stages as the result of multiple fragmentation of molecular cloud cores
(Bonnell \& Bate 1994, Burkert et al.\ 1997, Boss 1997, Klessen et al.\ 1998, Bate 2000). 
However, other scenarios
(e.g.\ stellar capture via close encounters, Turner et al.\ 1995) might lead to binary
formation as well. In addition, the fragmentation process may proceed with the coalescence
of the fragments.

The fact that circumstellar disks are an inevitable means to form stars
implies that the individual disks surrounding the binary components
represent tracers of the binary formation mechanism. The scattering
of emergent light from the star at disk surfaces and in the lobes of
disk envelopes leads to a net polarization (Bastien \& M${\rm \acute{e}}$nard 1990,
Whitney \& Hartmann 1992, Fischer et al.\ 1996). The position angle 
of the polarization vector is indicative for the disk orientation. The
formation of binaries by fission of a fragment will result in co-planar
individual disks (or a circumbinary disk) because of the conservation 
of angular momentum. Thus, the orientation of the polarization vector
for both components will be parallel. Otherwise, if binaries form
by capture, the disks will have a random orientation.
%(the time scale for
%disk alignment due to tidal forces is comparable to the disk lifetime)
Another interesting scenario is the interaction of protostellar disks
which would probably lead to hierarchical fragmentation where
in each cascade the number of disks is approximately doubled
(Watkins et al.\ 1998).

When addressing this question, we have to take into account 
that only the projection of the disk
onto the tangential plane can be measured. Therefore, in the case 
of an individual binary, we cannot rule out that
the disks have indeed different inclinations relative to the line of sight
although the position angles of the polarization of two components are
similar. However, a statistical study allows to test which
hypothesis (co-planarity, random orientation) is consistent with the
observed distribution of polarimetric position angles (see Sect.~\ref{proeff}). 

First attempts to address the issue of pre-main-sequence
binary polarization have been made 
(e.g., T Tau by Kobayashi et al.\ 1997, Ageorges et al.\ 1997, Fischer et al.\ 1998,
Monin et al.\ 1998, Jensen et al.\ 2000).
We briefly mention the results on Z\,CMa obtained by
speckle polarimetry (Fischer et al.\ 1998). In this case, the K band
linear polarization of the infrared primary amounts to 
$P_{\rm l} = 4.2\,\%\pm2.0\,\%, {\rm \gamma} = 173\pm34^{\rm o}$
($\gamma$ -- polarization angle) 
while the secondary (optical component) has 
$P_{\rm l} = 8.1\,\%\pm4.5\,\%, {\rm \gamma} = 102^{\rm o}\pm45^{\rm o}$.
This investigation revealed that the spatial orientation of the
individual disks is only marginally different, thus supporting the view of a
common origin. Monin et al.\ (1998) found four sources where the rotation
axes of both components are preferentially parallel but also one system
where the axes are clearly not parallel.
In a sample of 18 T Tauri binaries investigated by Jensen et al.\ (2000)
in the K band, approximately 70\,\% of the binaries have polarization angles 
being within $30^{\rm o}$ of each other.

We used the polarimetric mode of SOFI at the New Technology Telescope of ESO
to test two different scenarios of binary star formation.
For this purpose, a sample comprising 49 objects was selected from the
binary surveys of Reipurth \& Zinnecker (1993) and Ghez et al.\ (1997).

The near-infrared excesses of the targets mostly point to the presence 
of disks (e.g., Kenyon et al.\ 1996). 
For a major fraction of the sample, evidence for disks is also
obvious from other features (outflows, jets, Herbig-Haro objects).

The targets have angular separations $0.5\,\arcsec \le \rho \le 5.3\,\arcsec$.
Generally, the choice of the optimum wavelength for such an investigation 
will represent a compromise since a large net polarization is expected
at shorter wavelengths (due to the high scattering efficiency) but many 
of the targets are considerably reddened. 
%We conclude from the study of Ghez et al.\ (1997)
%that about 5\% of the seemingly single stars from the optical survey
%of Reipurth \& Zinnecker (1993) may have infrared companions.
%This would imply that these sources in fact are multiple systems. 

%-----------------------------------------------------------------------------------------
\section{The method}\label{method}

\subsection{The general idea}\label{tgidea}

The principal method that we apply in our investigation
has been firstly described by Monin et al.\ (1998).
It is based on the assumption that PMS stars are surrounded
by an optically thick disk being embedded in an optically thin
dust envelope. In the general case, the light emerging from such a configuration
is polarized whereby the degree of linear polarization $P_{\rm l}$
is strongly wavelength and inclination dependent and amounts to several 
percent in the K band.
While $P_{\rm l}$ is a function of many parameters (dust density
distribution in the disk, wavelength, orientation of the disk, etc.; see,
e.g., Fischer et al.\ 1996), the polarization angle $\gamma$ is only
a function of the orientation of the disk.
In the case of an optically thick disk and an optically thin circumstellar envelope,
it is identical with the orientation of the long semi-axis of the ellipse
which is the projection of the disk on the plane of the sky (in the case
of a disk seen face-on, $\gamma$ is not defined because the net polarization
is zero). The inclination of two circumstellar disks of a PMS binary projected
on the plane of the sky is therefore a function of the difference of the respective polarization
angles. The influence of the projection effect is described
in Sect.~\ref{proeff}.

%-------------------------------------------------------------------------------
\subsection{Polarization mechanisms}\label{pmsonp}

It is now well established that the optical and near-infrared polarization of light arising
from young stellar objects (YSO) is caused by light scattering
on dust grains. Assuming a centro-symmetric dust
density configuration around an illuminating star, the net polarization
arising from scattering by spherical dust grains
(see, e.g., Bastien \& M${\rm \acute{e}}$nard 1990,
Whitney \& Hartmann 1992, Fischer et al.\ 1996) or  - in a more realistic
scenario - by randomly oriented non-spherical grains (see Wolf et al.\ 2000),
is equal to zero. The polarization pattern is centro-symmetric.
If the star is surrounded by an optically thick disk
the net polarization will have a certain value:
\begin{enumerate}
\item The disk partially ``obscures'' the optically thin shell resp.\ 
  the centro-symmetric polarization pattern. The net polarization
  of the light arising from scattering in the shell is therefore not zero.
\item If the disk is not seen face-on, the scattering of light on the surface
  of the disk results in a non-centro-symmetric polarization pattern - 
  the resulting net polarization is not zero.
\end{enumerate}

The correlation between the polarization angle $\gamma$ and the orientation
of the disk is valid only under the following
assumptions:
\begin{enumerate}
\item The disk is optically thick at the wavelength of observation.
  Then the assumption of the obscuration effect is fulfilled.
\item The star + disk system is embedded in an optically thin shell.
  The light scattering by dust grains causes the polarization of light
  and backscattering on the disk. The latter results in a polarization
  angle being parallel to the long semi-axis of the projection
  of the disk surface on the plane of sky.
\end{enumerate}
Our binary sample was built using these constraints (see Sect.\ \ref{tsample}).

\begin{figure}
  \vspace*{18.5cm} 
  \includegraphics{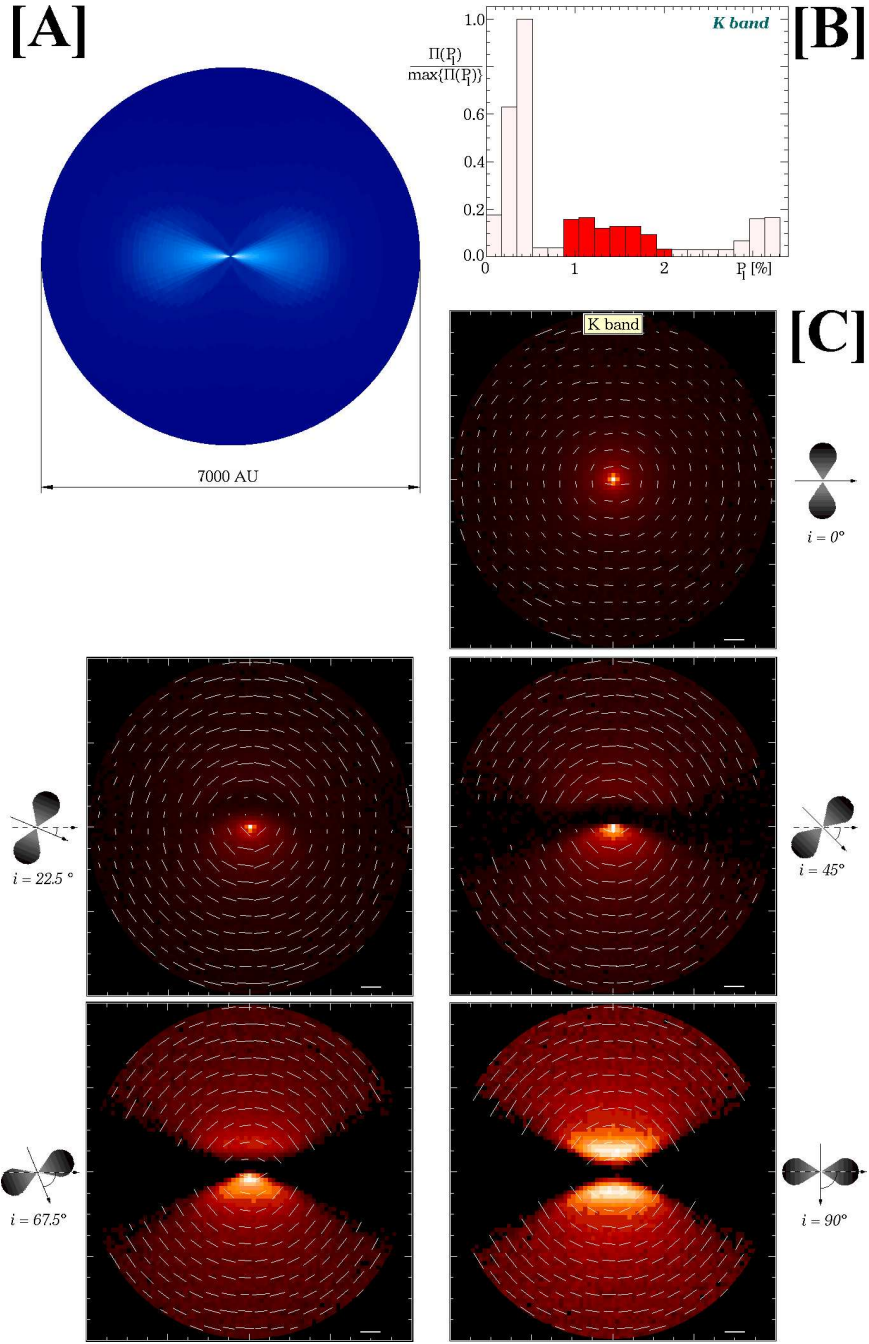}
  \caption{{\bf [A]} 
    Dust density distribution in the model of the circumstellar disk.
    {\bf [B]} 
    Histogram of the K band net polarization arising from the YSO configuration.
    {\bf [C]}
    Intensity maps with overlayed polarization pattern of the disk
    for different inclinations in the K band. The polarization scale in the lower
    right edges of the images symbolizes a polarization degree of 100\%.}\label{simuyso}
\end{figure}
In Fig.~\ref{simuyso} polarization maps and the resulting probability distribution 
of the observable net polarization in the K band for a model of a low-mass YSO 
being surrounded by a circumstellar disk are shown.
The spatial density structure (see Fig.~\ref{simuyso}[A]) and initial temperature distribution 
of the circumstellar disk ($M=2.3\cdot10^{-3}$\,$M_{\sun}$) results from hydrodynamical simulations
performed by Yorke (1999, priv.\ comm.). 
The hydrodynamic code solves the standard equations of hydrodynamics 
with radiation transport and the Poisson equation for the gravitational potential 
(Black \& Bodenheimer 1975).
The two-dimensional hydrodynamic code of R${\rm {\grave o} {\dot z}}$yczka (1985)
with second-order accurate advection is employed. Shocks are treated
by including artificial viscosity. Physical viscosity is not included and
angular momentum transport during the collapse is assumed to be
negligible.
To derive the polarization $P_{\rm l}(i)$ of this system as a function of the inclination
$i$ perpendicular to the plane of sky, we simulated the radiative transfer (RT) with 
a three-dimensional continuum RT code which is based on the Monte-Carlo method
(see, e.g., Wolf et al.\ 1999).
In addition to the results from the hydrodynamical simulations, we introduce
the following RT parameters:
spherical dust grains consisting of ``astronomical'' silicates 
(optical data from Draine \& Lee 1984, radius 0.12\,${\rm \mu}$m); 
star: effective temperature $T_{\rm eff}=6000$\,K, $L = 2\,L_{\sun}$; 
wavelength range for the simulation of the radiative transfer:
 $0.03 \ldots 2000\,{\rm \mu}$m.

%-----------------------------------------------------------------------------------------
\subsection{Inclination of disks - Projection effect}\label{proeff}

\begin{figure} 
  \vspace*{7.0cm} 
  \includegraphics{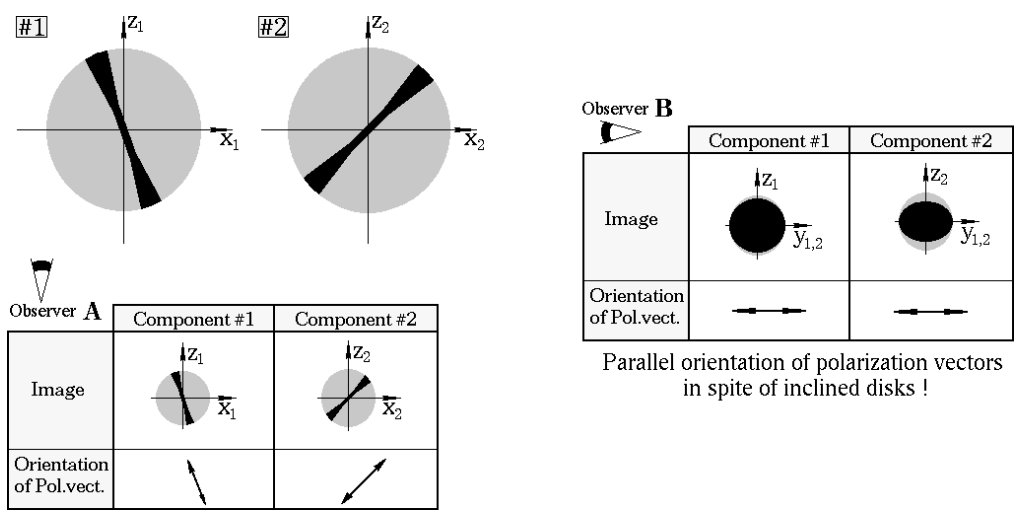}
  \caption{Explanation of the projection effect.
    In the upper left part of the figure two flared disks (seen edge-on, black)
    embedded in an optically thin circumstellar shell (grey)
    are shown. The y-axis of the cartesian coordinate systems
    [$(x_{\rm 1},y_{\rm 1},z_{\rm 1})$;$(x_{\rm 2},y_{\rm 2},z_{\rm 2})$]
    is oriented perpendicular to the x-z plane pointing into the paper plane.
    If the plane of the sky is the
    $x_{\rm 1}-z_{\rm 1}$ (resp.\ the $x_{\rm 2}-z_{\rm 2}$) plane,
    the real inclination of the disks against each other can be measured
    (Observer A; see lower left table). If the 
    $y_{\rm 1}-z_{\rm 1}$ (resp.\ the $y_{\rm 2}-z_{\rm 2}$) plane
    is the plane of the sky (Observer B), the polarization angle
    $\gamma$ of both disks is parallel to the according y-axis.
    Therefore - due to the projection effect -, these disks would be assumed
    to be co-planar. Only on the basis of a large binary
    sample, a decision whether the individual disks of the binary components are in general
    co-planar or oriented randomly can be made (see Sect.~\ref{proeff}).}\label{diskpro}
\end{figure}

One has always to take into account that only the orientation of
the disk resp.\ the orientation of its spin axis
projected onto the plane of the sky can be derived from the
linear polarization orientation angle $\gamma$.
If the spin axis is rotated in the plane of the sky,
the polarization angle $\gamma$ rotates accordingly and the net polarization
remains unchanged.
In contrast to this, the polarization angle $\gamma$ remains
unchanged if the spin axis is rotated perpendicular to the plane
of the sky. In the second case the net polarization $P_{\rm l}$
changes. Because the dust density structure of the individual disks
is unknown, the inclination of a disk perpendicular to the plane
of sky cannot be derived from $P_{\rm l}$.
As a consequence of this situation, the real angle between the spin axes of two disks
$\alpha$ {\em cannot} be derived with this technique.
For example, a measured position angle difference $\Delta\gamma=0\deg$
is not a reliable indicator for parallel spin axes (see Fig.~\ref{diskpro}).

However, this problem can be solved considering a large binary sample.
If in each binary the disks are randomly oriented,
the probability $\Pi(\Delta\gamma)$ for a pair of disks to have an inclination
projected on the plane of sky amounting to $\Delta\gamma$ is constant.
Even if the correct inclinations (which cannot be measured with this technique)
differ from the measured inclinations in the plane of the sky,
the probability distribution $\Pi(\Delta\gamma)$ - based on measurement
of the polarization angles of the binary components - is {\sl not}
modified in the case of randomly oriented disks.
This comes from the fact that for every orientation of the disk
in the plane of the sky (represented by the polarization angle $\gamma$)
the probability for the disk to be inclined by a certain angle
perpendicular to the plane of the sky is constant.

\begin{figure} 
  \vspace*{4.5cm} 
  \includegraphics{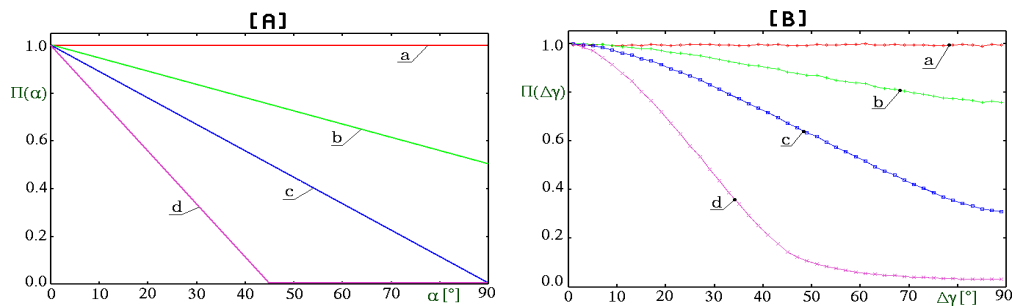}
  \caption{Correlation between {\bf [A]} the distribution of the real inclinations $\Pi(\alpha)$
    of the disks and {\bf [B]} the distribution of the corresponding
    position angle differences $\Pi(\Delta\gamma)$ (see Sect.~\ref{proeff}).
    {\bf (a)}: perfect random orientation of the disks;
    {\bf (b)$\ldots$(d)}: increasing alignment of the disks.}\label{simuinc}
\end{figure}
In Fig.~\ref{simuinc} the correlation between the real inclination $\alpha$
of the disks and the position angle difference $\Delta \gamma$
is shown for different probability distribution functions $\Pi(\alpha)$\footnote{These
results are based on Monte-Carlo simulations. Each distribution $\Pi(\Delta \gamma)$ 
was determined based on the projection of $10^7$ pairs of disks - following 
the distribution $\Pi(\alpha)$ - onto the plane of the sky.}.
For perfectly randomly oriented disks ($\Pi(\alpha)$=const.), 
we find $\Pi(\Delta\gamma)$=const.\,.
An alignment of disks - being characterized by a decrease of the probability
distribution function $\Pi(\alpha)$ for increasing $\alpha$ - results
in a decrease of the measured distribution $\Pi(\Delta\gamma)$.
One has also to be aware, that even in the case of strongly
- but not perfectly - aligned disks for which $\Pi(\alpha)$=0 for 
$\alpha>\alpha_{\rm 0}$ ($\alpha_{\rm 0}>0$; see Fig.~\ref{simuinc}, distribution {\em d}),
there is still a probability to measure position angle differences $\Delta\gamma>\alpha_{\rm 0}$.

%-----------------------------------------------------------------------------------------
\subsection{Influence of the interstellar polarization}\label{infismsc}

As discovered by Hiltner (1949), the interaction of light of distant stars
with the interstellar medium (ISM) leads to a significant linear polarization.
Because this additional ``polarizer'' (polarization degree $P_{\rm l,ISM}$,
orientation angle $\gamma_{\rm ISM}$) influences the observed light of both binary components,
the observed position angle difference $\Delta\gamma$ may be reduced.
Thus, even in the case of perfectly randomly oriented disks, this may lead
to an apparent alignment of the disks because the probability distribution
function $\Pi(\Delta\gamma)$ would increase towards decreasing $\Delta\gamma$.

\begin{figure}
  \vspace*{7.0cm} 
  \includegraphics{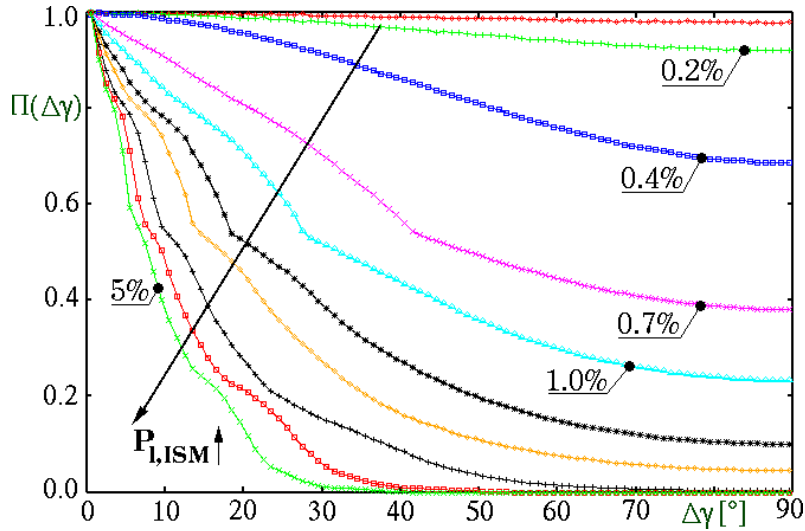}
  \caption{Probability distribution function $\Pi(\Delta\gamma)$ for randomly 
    oriented disks as a function of the degree of linear polarization of the 
    ISM $P_{\rm l, ISM}$. The decrease of $\Pi(\Delta\gamma)$ steepens with
    increasing polarization $P_{\rm l, ISM}$.
    $P_{\rm l, ISM}$= 0.1, 0.2, 0.4, 0.7, 1.0, 1.5, 2.0, 3.0, 4.0, and 5.0\,\%\,.}\label{simuism}
\end{figure}
To estimate the influence of the interstellar polarization (ISP) on our results,
we derived the K-band polarization of the model of a low-mass YSO described 
in Sect.~\ref{pmsonp}. The second step was to combine two of these disk under 
the assumption of random pairing  ($\Pi(\alpha)$=const.). 
Finally, the measurable apparent inclination of these disks
$\Delta\gamma$ in the plane of sky was determined taking into account the additional polarization
by the ISM. Based on $10^7$ pairs of disks, the resulting distribution function
$\Pi(\Delta\gamma)$ as a function of $P_{\rm l,ISM}$ is shown in Fig.~\ref{simuism}.
The decrease of the distribution $\Pi(\Delta\gamma)$ towards increasing $\Delta\gamma$
is strengthened by the increase of $P_{\rm l, ISM}$.

%-----------------------------------------------------------------------------------------
\section{Observations}\label{obs}

\subsection{The sample}\label{tsample}

As pointed out in Sect.\ \ref{pmsonp}, we have to consider PMS stars in an 
evolutionary stage in which they are still surrounded by an optically thick disk being
embedded in an optically thin shell.
The objects fulfilling these criteria are classical T Tauri
stars. Therefore our sample consists of 49 close binaries 
(separations: 0.5\arcsec$\ldots$5.3\arcsec) from the surveys 
of Reipurth \& Zinnecker (1993), Ghez et al.\ (1997), and Prato \& Simon (1997).

%---------------------------------------------------------------------------
\subsection{Observation strategy}\label{obsstrat}

The polarization data were obtained in 1999 March 2-4
and May 28-30 at the European Southern Observatory (ESO).
The images were taken using SOFI in imaging polarization
mode at the New Technology Telescope (NTT). The pixel scale in this mode
is 0.292\arcsec\ per pixel. 

With SOFI, polarimetry is peformed by inserting in the parallel beam
a Wollaston prism which splits the increasing light rays into 2 orthogonally
polarized beams separated by 48\arcsec. 
%Every object
%in the field has therefore two images on the CCD detector.
%In order to avoid any overlapping of different images and to reduce the
%sky contribution, an aperture mask is put at the focal plane of the
%telescope. 
For the derivation of the Stokes vector components Q and U,
the objects were observed at two different orientations of the
Wollaston prism differing by 45 degrees.

To increase the signal-to-noise ratio (SNR) and for a sufficient badpixel
and background subtraction we observed each object at each
orientation of the Wollaston prism 5 times (offset: 25\arcsec).
%The offset
%between 2 consecutive exposures is 25\arcsec\ and is therefore
%much larger than the separation of two binary components.

%-----------------------------------------------------------------------------------------
\section{Results}\label{results}

\begin{figure}
  \vspace*{5.0cm} 
  \includegraphics{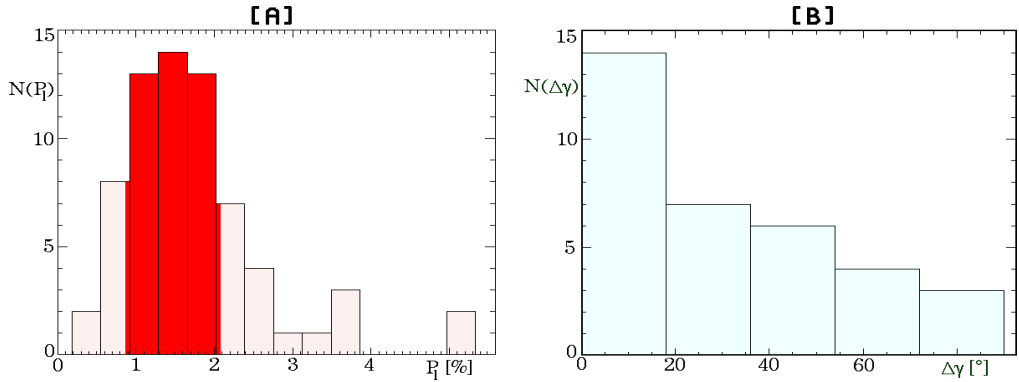}
  \caption{{\bf [A]} Histogram of the polarization measured for 2$\times$34 binary
    components. The dark grey region markes the ranges in which we found a local maximum
    of the linear polarization in out model simulations (for comparison see Fig.~\ref{simuyso}[B]).
    {\bf [B]} Distribution of the position angle differences $\Delta\gamma$ 
    of the disks in 34 binaries. For comparison see Fig.~\ref{simuinc} 
    and Fig.~\ref{simuism}.}\label{resultx}
\end{figure}
The histograms of the linear polarizations and the position angle differences
$\Delta\gamma$ based on 34 binaries are shown in Fig.~\ref{resultx}.
The remaining objects were excluded because the error of polarization
and therefore the error of the orientation angle $\gamma$ (for the formalism of
error estimation see, e.g., Wardle \& Kronberg 1974, di Serego Alighieri 1998) 
of at least one component of the binary system
was too large ($P_{\rm l}/\sigma(P_{\rm l})<3$). Thus, preferentially those binaries
in which at least one component shows a very low linear polarization
($\le 1$\%), were excluded. This can also be seen in the histogram for $P_{\rm l}$
which is otherwise (for $P_{\rm l}>1$\%) in very good agreement with the simulated
histogram (see Sect.~\ref{infismsc}, Fig.~\ref{simuyso}[B]).
The distribution ($N(\Delta\gamma)$) shows a strong
decrease towards large angles $\Delta\gamma$. Assuming the 
ISP to cause this behaviour, the degree of linear polarization had to exceed 1\%
(for comparison see Fig.~\ref{simuism}[B]).
In contrast to this, the mean degree of linear polarization caused by the ISM
in the K band was found to be well below 1\,\% in average (see, e.g., Nagata 1990).
Thus, the distribution $\Pi(\Delta\gamma)$ reflects the intrinsic polarization
of the binary components and therefore the alignment of their disks.

%-----------------------------------------------------------------------------------------
\section{Conclusions}\label{concl}

Based on K band observations of 49 classical T Tauri binary stars, we found
intrinsic linear polarization for both components in 34 binary systems.
The polarization can be explained by light scattering on spherical dust grains
under the assumption of the presence of a circumstellar disk. We performed
self-consistent RT simulations for a low-mass YSO surrounded by a circumstellar
disk with a structure resulting from hydrodynamical simulations. The probability distribution of
the polarization degree fits very well to the observed distribution derived from
the polarization of all binary components.

We showed that real inclinations of the circumstellar disks in single binary systems
cannot be measured using the position angle difference $\Delta\gamma$.
But it turned out that, based on a large binary sample, one can decide if there
exists an alignment mechanism. The measured probability distribution function
$\Pi(\Delta\gamma)$ shows a steep decrease towards large inclinations $\Delta\gamma$.
Based on RT simulations we showed that this behaviour cannot be explained by
the influence of the ISP.

We can therefore conclude that the disks in our sample are preferentially aligned.
Thus, if stellar capture plays a role during the binary formation process,
it must be followed by disk alignment processes on a timescale much smaller
than the lifetime of the circumstellar disks. 
Otherwise, fragmentation as the basic binary formation process
could explain the observed inclination distribution without any assumptions.
The tail towards large inclinations can be explained by the projection effect
(see Sect.~\ref{proeff}). Moreover, turbulence in the molecular cloud during 
the fragmentation process might lead to misaligned fragments as well 
(R. Klein, priv.\ comm.).

%-----------------------------------------------------------------------------------------
\acknowledgments

We wish to thank H.W.\ Yorke for providing the spatial density and temperature
distribution of a protostellar disk.
This research was supported by the DFG grant Ste 605/10 within the program
``Physics of star formation''. 
It is based on observations collected at the European Southern Observatory.
%It has made use of NASA's Astrophysics Data System Bibliographic Services.

%-----------------------------------------------------------------------------------------


\begin{references}
\reference Ageorges N., Eckart A., Monin J.-L., M$\rm \acute{e}$nard F., A\&A 326, 632
\reference Bastien P. \& M${\rm \acute{e}}$nard F., 1990, ApJ 364, 232
\reference Bate M.R., 2000, MNRAS 314, 33
\reference Black D.C. \& Bodenheimer P., 1975, ApJ 199, 619
\reference Bouvier J., Rigaut F., Nadeau D., 1997, A\&A 323, 139
\reference Bonnell I.A. \& Bate M.R., 1994, MNRAS 271, 999
\reference Boss A.P., 1997, ApJ 483, 309
\reference Burkert A., Bate M.R., Bodenheimer P., 1997, MNRAS 289, 497
\reference Draine B.T., Lee H.M., 1984, ApJ 285, 89
\reference Fischer O., Henning Th., \& Yorke H.W., 1996, A\&A 308, 863
\reference Fischer O., Stecklum B., \& Leinert Ch., 1998, A\&A 334, 969
\reference Ghez A.M., McCarthy D.W., Patience J.L., Beck T.L., 1997, ApJ 481, 378
\reference Hiltner W.A., 1949, ApJ 109, 471
\reference Jensen E., Donar A.X., Mathieu R.D., 2000, in {\sl Birth and Evolution of Binary stars},
  Ed. B. Reipurth \& H. Zinnecker
\reference Kenyon S.J., Yi. I. \& Hartmann L., 1996, ApJ 462, 439
\reference Kobayashi N., Nagata T., Tamura, M., et al., 1997, ApJ 481, 936
\reference Klessen R.S., Burkert A., Bate M.R., 1998, ApJL 510, L205
\reference K\"ohler R., Leinert Ch., 1998, A\&A 331, 977
\reference Leinert Ch., Richichi A., Haas M., 1997, A\&A 318, 472
\reference Monin J.-L., M${\rm \grave{e}}$nard F., Duch${\rm \hat{e}}$ne G., 1998, A\&A 339, 113
\reference Nagata T., 1990, ApJ 348, L13
\reference Prato L., Simon M., 1997, ApJ 474, 455
\reference Reipurth B. \& Zinnecker H., 1993, A\&A 278, 81
\reference R${\rm {\grave o} {\dot z}}$yczka M., 1985, A\&A 143, 59
\reference di Serego Alighieri S., 1998, in {\sl Instrumentation for Large Telescopes},
  Ed. J.M. Rodriguez, Cambridge University Press
\reference Simon M., Ghez A.M., Leinert Ch., et al., 1995, ApJ 443, 625
\reference Turner J.A., Chapman S.J., Bhattal A.S., et al., 1995, MNRAS 277,705
\reference Whitney B.A., Hartmann L., 1992, ApJ 395, 529
\reference Wardle J.F.C., Kronberg P.P., 1974, ApJ 194, 249
\reference Watkins S.J., Boffin H.M.J., Francis N., Whitworth A.P., 1998, ASP 132, 430
%\reference Wolf S., Henning Th., 2000, Comp.\ Phys.\ Comm., in press
\reference Wolf S., Henning Th., Stecklum, B., 1999, A\&A 349, 839
\reference Wolf S., Voshchinnikov N., Henning Th., 2000, in preparation
\end{references}
\end{document}